\documentstyle[12pt,a4wide]{article}
\begin{document}

\author{B. Linet \thanks{E-mail: linet@celfi.phys.univ-tours.fr} \\
\small Laboratoire de Math\'ematiques et Physique Th\'eorique \\
\small CNRS/UPRES-A 6083, Universit\'e Fran\c{c}ois Rabelais \\
\small Parc de Grandmont 37200 TOURS, France}
\title{\bf Entropy bound for a charged object \\
from the Kerr-Newman black hole}
\date{}
\maketitle

\begin{abstract}

We derive again the upper entropy bound for a charged object by 
employing thermodynamics of the Kerr-Newman black hole linearised with 
respect to its electric charge.


\end{abstract}

\thispagestyle{empty}

Recently, Bekenstein and Mayo \cite{bek1} and Hod \cite{hod} have obtained 
an upper entropy bound for a charged object by requiring the validity of 
thermodynamics of the Reissner-Nordstr\"{o}m black hole linearised with 
respect to its electric charge. They have found again the proposal of 
Zaslaskii \cite{zas} derived in another context.
The proof takes into account the general existence of an electrostatic 
self-force acting on a point charge in a gravitational field 
\cite{dew,unr,vil} which has been exactly determined for the Schwarzschild 
black hole \cite{smi,zel,lea1}.

A question occurs immediately: what is the upper entropy bound for a charged
object by requiring the validity of thermodynamics of the Kerr-Newman 
black hole linearised with respect to its electric charge ? It is possible
to answer because the electrostatic self-force for a point charge on the
symmetry axis of the Kerr black hole has been previously determined 
\cite{loy,lea2}. The purpose of this work is to derive again the entropy bound
in this situation according to the method initially due to Bekenstein 
\cite{bek2} for a neutral object falling in a Schwarzschild black hole.

The Kerr black hole is characterized by the mass $m$ and the angular momentum 
per unit mass $a$ satisfying $m^2>a^2$. In the coordinate system 
$(t,r,\theta ,\varphi )$, the Kerr metric is
\begin{eqnarray}\label{K}
\nonumber ds^2&=& \left( 1-\frac{2mr}{\Sigma}\right) dt^2-\frac{\Sigma}
{\triangle}dr^2-
\Sigma d\theta^2+\frac{4amr\sin^2\theta}{\Sigma}dtd\varphi \\
& & -\sin^2\theta \left( r^2+a^2+\frac{2a^2mr\sin^2\theta}{\Sigma}\right)
d\varphi^2
\end{eqnarray}
with $\triangle =r^2-2mr+a^2$ and $\Sigma =r^2+a^2\cos^2\theta$.
The Kerr-Newman black hole linearised with respect to its electric charge $q$
is described by metric (\ref{K}) plus an electromagnetic test field 
having the components
\begin{equation}\label{A}
A_t=\frac{qr}{\Sigma}\quad A_r=0\quad A_{\theta}=0 \quad
A_{\varphi}=-\frac{qar\sin^2\theta}{\Sigma} .
\end{equation}
The area of the Kerr-Newman black hole has the expression
\begin{equation}\label{S}
{\cal A}(m,a,q)=4\pi \left[ \left(m+\sqrt{m^2-a^2-q^2}\right)^2+a^2\right]
\end{equation}
which reduces to
\begin{equation}\label{SL}
{\cal A}(m,a,q)\approx 4\pi\left[ 2m^2+2m\sqrt{m^2-a^2}-
\frac{q^2m}{\sqrt{m^2-a^2}}-q^2 \right]
\end{equation}
for a Kerr-Newman black hole linearised with respect to its electric charge $q$.
In thermodynamics of the black hole, the entropy $S_{BH}$ is given by
\begin{equation}\label{SBH}
S_{BH}(m,j,q)=\frac{1}{4}{\cal A}(m,j/m,q) .
\end{equation}
in terms of the thermodynamical variables $m$, $j$ and $q$.

On the symmetry axis of metric (\ref{K}),
outside the outer horizon $r_+=m+\sqrt{m^2-a^2}$, we
consider a charged object with a mass $\mu$, an electric charge $e$, 
an entropy $S$
and a radius $R$ whose the own gravitational field is negligeable and
the electromagnetic field generated by the charge $e$ is a test field.
By making use of a quasi-static assumption, we restrict ourselves to the
case where the charged object is at rest. We suppose that its total energy
${\cal E}$ with respect to a stationary observer at infinity coincides with
the one of a massive point charge located at $r=r_0$ and $\theta =0$ with
$r_0>r_+$. Obviously, the dimension $R$ of this object is taken as the 
proper length along the symmetry axis of metric (\ref{K}). This proper length 
$\ell$ from the outer horizon to the position $r_0$ for $\theta =0$
has the expression
\begin{equation}\label{L}
\ell =\int_{r_+}^{r_0}\frac{\sqrt{r^2+a^2}}{\sqrt{r^2-2mr+a^2}}dr .
\end{equation}

The total energy ${\cal E}$ of a massive point charge is the sum of the 
energy $W_{mass}$ of the mass $\mu$, the electrostatic energy $W_{ext}$ of
the charge $e$ in the exterior electromagnetic field (\ref{A}) and the
electrostatic self-energy $W_{self}$ of the charge $e$. The electrostatic 
self-force $f_{self}^{i}$ exerted on the point charge by its self-field 
has the expression \cite{loy,lea2}
\begin{equation}\label{force}
f_{self}^{i}=\frac{e^2mr_0}{(r_{0}^{2}+a^2)^2}\delta_{1}^{i}
\end{equation}
in Fermi coordinates at the position of the point charge. We can easily 
determined $W_{self}$ so that it yields self-force (\ref{force}); we find
\begin{equation}\label{self}
W_{self}=\frac{1}{2}\frac{e^2m}{r_{0}^{2}+a^2} .
\end{equation}
So, the total energy ${\cal E}$ is given by
\begin{equation}\label{E}
{\cal E}=\frac{\mu\sqrt{r_{0}^{2}-2mr_{0}+a^2}}{\sqrt{r_{0}^{2}+a^2}}+
\frac{eqr_0}{r_{0}^{2}+a^2}+\frac{e^2m}{2(r_{0}^{2}+a^2)} .
\end{equation}
For the charged object, its last state which is possible 
outside the outer horizon is defined by $\ell =R$. By assuming that 
$\ell$ is small, we deduce from (\ref{L}) the asymptotic form
\begin{equation}\label{LA}
\ell \sim \frac{\sqrt{2}(r_{+}^{2}+a^2)}{(m^2-a^2)^{1/4}}\sqrt{r_0-r_+}
\quad {\rm as}\quad r_0 \rightarrow r_+ .
\end{equation}
By substituting (\ref{LA}) into (\ref{E}), we obtain the energy 
${\cal E}_{last}$ of the last state 
\begin{equation}\label{last}
{\cal E}_{last}\sim \frac{\mu R \sqrt{m^2-a^2}}{r_{+}^{2}+a^2}+
\frac{eqr_+}{r_{+}^{2}+a^2}+\frac{e^2m}{2(r_{+}^{2}+a^2)}\quad {\rm as}
\quad R\rightarrow 0 .
\end{equation}

We are now in a position to apply thermodynamics of the Kerr-Newman black
hole when the charged object falls infinitely slowly along the symmetry axis
until the absorption inside the outer horizon which is the final state.
This is the original method of Bekenstein \cite{bek2}
for a neutral object in the Schwarzschild black hole. The final state is 
again a Kerr-Newman black hole but with the new parameters
\begin{equation}\label{p}
m_f=m+{\cal E}_{last} \quad , \quad  j_f=j \quad {\rm and}\quad q_f=q+e .
\end{equation}
In the last state outside the outer horizon, the total entropy is 
$S_{BH}(m,j,q)+S$ but after the absorption 
the entropy is only $S_{BH}(m_f,j_f,q_f)$ in the final 
state. By virtue of the generalised 
second law of thermodynamics, we must have
\begin{equation}\label{i}
S_{BH}(m_f,j_f,q_f)\geq S_{BH}(m,j,q)+S .
\end{equation}
We can calculate $\triangle S_{BH}=S_{BH}(m_f,j_f,q_f)-S_{BH}(m,j,q)$ from 
expression (\ref{SL}) with increments (\ref{p}) by keeping only linear
terms in ${\cal E}_{last}$. We find
\begin{equation}\label{AP}
\triangle S_{BH}=\frac{2\pi}{\sqrt{m^4-j^2}}\left[ 2m\left( m^2+
\sqrt{m^4-j^2}\right){\cal E}_{last}- \left( m^2+\sqrt{m^4-j^2}\right) 
\left( eq+\frac{1}{2}e^2\right) \right]
\end{equation}
that we can rewrite under the form 
\begin{equation}\label{delta}
\triangle S_{BH}=\frac{2\pi}{\sqrt{m^2-a^2}}\left[ (r_{+}^{2}+a^2)
{\cal E}_{last}-eqr_+-\frac{1}{2}e^2r_+ \right] .
\end{equation}
We simplify expression (\ref{delta}) by using ${\cal E}_{last}$ given by
(\ref{last}). Then, taking into account inequality (\ref{i}), we thus obtain
the desired entropy bound
\begin{equation}\label{bound}
S\leq 2\pi \left( \mu R-\frac{1}{2}e^2\right) . 
\end{equation}

In conclusion, we have extended the works of Bekenstein and Mayo \cite{bek1}
and Hod \cite{hod}
by employing thermodynamics of the Kerr-Newman black hole instead of the 
Reissner-Nordstr\"{o}m black hole. These kinds of method for determining the
entropy bound is without pretending to any rigour. However, they confirm the
physical importance of the electrostatic self-force acting on a point charge 
in a background black hole although its physical relation with the entropy 
of a charged object is not clear.

\end{document}